\documentclass{PoS}

\title{SUSY scenarios according to EWSB}

\ShortTitle{SUSY scenarios according to EWSB}

\author{\speaker{Radovan Derm\' \i\v sek}%
       \\ \\
       Physics Department, Indiana University, Bloomington, IN 47405, USA\\
       E-mail: \email{dermisek@indiana.edu}}

\abstract{This talk provides a limited review of SUSY scenarios with the focus on the way electroweak symmetry breaking is achieved and understood under different assumptions. Various aspects of naturalness and their implications are discussed and compared. }

\FullConference{The European Physical Society Conference on High Energy Physics\\
                 5-12 July\\
                 Venice, Italy}

\begin{document}

\section{Introduction: EWSB in the MSSM and a simple naturalness argument}

In supersymmetric (SUSY) models, electroweak symmetry breaking (EWSB) can be achieved radiatively. The relevant combination of parameters is given by the supersymmetric Higgs mass, also known as the $\mu$- term, and soft SUSY breaking mass squared of the Higgs doublet that couples to the top quark, $M_Z^2 \simeq -2\mu^2(M_Z) -2m_{H_u}^2(M_Z)$, with the latter being driven to negative values in vast ranges of the parameter space by masses of scalar tops and gluino. This is certainly a very nice feature with an immediate consequence  that the EW scale is related to superpartner masses.

For given $\tan \beta$ (which sets values of Yukawa couplings), we can solve the RG equations and write the same formula in terms of boundary conditions at the scale where soft masses are generated. For the MSSM with boundary conditions at the GUT scale and $\tan \beta = 10$
 we have:
\begin{equation}
M_Z^2 \simeq -1.9\mu^2 + 5.9 M_3^2 -1.2 m_{H_u}^2 +1.5 m_{\tilde t}^2 -0.8A_t M_3 + 0.2 A_t^2 +\dots , 
\label{eq:MZ}
\end{equation}
where $M_3$ is the gluino mass (at the GUT scale), $A_t$ is the top soft 
SUSY breaking trilinear coupling, and for simplicity we have defined $m_{\tilde t}^2 \equiv (m_{\tilde t_L}^2 + m_{\tilde t_R}^2)/2$
and assumed that  $m_{\tilde t_L} \simeq m_{\tilde t_R}$ . Other soft masses appear with much smaller  coefficients and we neglect them in our discussion. The coefficients
in this expression depend only on $\tan \beta$ (they do not change dramatically when varying $\tan \beta$ between 5 and 50) and $\ln (M_{G}/M_Z)$.

We see that the same parameters that determine masses of superpartners %, namely stop and gluino and chargino, 
appear with order one coefficients in the formula for the EW scale.  
This leads to the expectation that superpartners should be ``near''  the EW scale. However, how near depends on something that is just in our minds; it depends on what we consider reasonable and natural. Just as with any esthetic argument even naturalness depends on the point of view and evolves with time.  

The most common and simplest naturalness argument is 
based on the largest individual contribution to the EW scale and is typically estimated by a variety of probabilistic methods or sensitivity measures. Requiring that no individual contribution in Eq.~(\ref{eq:MZ})   is significantly larger than $M_Z^2$ leads to the expectation that stops, gluino and chargino should be lighter than few hundred GeV. However, this expectation from our simple naturalness argument seems to contradict direct limits: $\gtrsim 2$ TeV for the gluino mass, $\gtrsim 1$ TeV for stop masses, $\gtrsim 1$ TeV for the chargino mass (we should keep in mind that these limits depend on many assumptions and are not universal). There are also constraints on stop masses from the Higgs boson mass. These depend on the mixing in the stop sector and, in the MSSM, require stops above 1 TeV for maximal mixing and extend to $\sim 10$ TeV for no mixing.

The limits on superpartners suggest  that the left hand side of Eq.~(\ref{eq:MZ})  is $\sim 1000$ times smaller than some terms on the right hand side.
Clearly, if the contribution of one parameter is 1000 times larger than the desired value, there has to be something else that cancels this contribution very precisely, with three digit precision. In other words, keeping all the remaining parameters fixed, the most contributing parameter needs to be tuned at 0.1\% level. 

This is certainly a very puzzling situation. If SUSY is really there then it means that something from our discussion so far is not quite right. The possibilities can be roughly organized into three groups and examples of each will be discussed in the following sections. The first possibility is that  the EWSB is simply  fine tuned or the SUSY breaking mechanism provides special relations between parameters that lead to seemingly fine tuned scenario, or  the limits on superpartners are somehow avoided.\footnote{There is extensive literature that fits to this category with respect to both models and searches. Detailed  discussion of this possibility is beyond the scope of this limited review. I refer the reader to the current status of SUSY searches~\cite{SUSY-status}, few reviews that discuss SUSY scenarios in connection with  naturalness~\cite{Bechtle:2015nta, Feng:2013pwa, Dermisek:2009si} and any other recent review on the status of supersymmetry.}
The second possibility is that the MSSM is not the right model in which case the relevant superpartners can contribute differently to the EW scale compared to what is expected in the MSSM. Models where gluino~\cite{Fox:2002bu}, chargino~\cite{Cohen:2015ala, Nelson:2015cea} and stop~\cite{Dermisek:2016tzw} contribute significantly less than in the MSSM exist. I will briefly discuss the model with  stop having different properties. 
Finally, the third possibility is that the naturalness argument based on the largest individual contribution to the EW scale is too strong requirement.

\section{SUSY is tuned or special}

One of the simplest interpretation of current experimental situation is that SUSY is heavy and the model parameters are properly tuned to obtain the EW scale significantly below the scale of superpartners. This can be just accidental or it can be motivated by anthropic reasoning or by other statistical arguments. If this is the case then the only solid indication for superpartner masses comes from the Higgs boson mass. 

In the MSSM, for $\tan \beta $ larger than few, the stop masses needed to obtain the measured value of the Higgs boson mass are: $m_{\tilde t} \gtrsim 7$ TeV for $X_t/m_{\tilde t} = 0$, $m_{\tilde t} \gtrsim 3$ TeV for $X_t/m_{\tilde t} \simeq -1$ and $m_{\tilde t} \gtrsim 1$ TeV for $X_t/m_{\tilde t} \simeq -2$, where $X_t/m_{\tilde t}$  specifies the mixing in the stop sector.  The required stop masses depend only mildly on other SUSY parameters. As $\tan \beta $ approaches 1 the needed stop masses are pushed up by several orders of magnitude.   
Thus, if fine tuning is not an issue for whatever reason, the experiments only started touching the suggested range.

There is also a possibility that SUSY is not fine tuned but rather somewhat special. A simple way to improve on naturalness is to significantly lower the mediation scale. The coefficients appearing in front of soft masses in Eq.~(\ref{eq:MZ}) depend mostly on the top Yukawa coupling, $\lambda_t$, strong coupling constant, $\alpha_3$, and  the mediation scale, $\Lambda$. All the coefficients, except for those multiplying $\mu^2$ and $m_{H_u}^2$, go to zero as the mediation scale decreases. Thus the gluino and stops can in principle decouple from the EW scale. However, in the limit when we completely solve the problem we also lose the understanding of EWSB (gluino and stops drive the Higgs soft mass squared to negative values). Nevertheless, one can definitely improve on naturalness significantly and we will return to a lower mediation scale in the next section. 

Another possibility is that a SUSY breaking model provides relations between parameters that are just such that contributions from different  terms to large extent cancel. It is possible to imagine that in a complete model the combined right hand side of Eq.~(\ref{eq:MZ}) would be much smaller than individual terms. One should keep in mind however, that the coefficients of various terms,  and thus the needed relation,  are  $f(\lambda_t,\alpha_3,\Lambda)$  and it seems like a huge coincidence that SUSY breaking scenario would know about all this. 

Yet another possibility is that 
SUSY is in an island of the parameter space where searches are not sensitive.
Or, the scenario predicts unsearched for or hard to search for decay modes, or one can imagine a scenario where superpartners have complicated or too many decay modes. However,  if SUSY is light and somehow escaped detection, an extra mechanism has to be supplied to increase the Higgs boson mass.

\section{Beyond the MSSM: Stops are little different}

As already mentioned, lowering the mediation scale is certainly a possible way to alleviate fine tuning to some degree. It works really well for the gluino contribution to $m_{H_u}^2$. Since it is a two-loop effect, even a 2 TeV gluino does not result in more than 10\% tuning in EWSB as far as the mediation scale is below $\sim $ 1000 TeV.
However, $\sim $10 TeV stops, suggested by the Higgs mass assuming no mixing in the stop sector, feed to  $m_{H_u}^2$  very fast. Just one decade of RG running results in $\sim (3 \; {\rm TeV})^2$ contribution to   $m_{H_u}^2$       indicating $\sim 0.1\%$ tuning.
In the MSSM this contribution is either there or one must give up the radiatively driven EWSB as a robust feature (models can be constructed where EWSB happens by design and radiative  corrections are not essential). 

In extensions of the MSSM with vector-like quarks (or complete vector-like families) the stop can be a mixture of a state that couples to the Higgs boson but does not have a large soft mass and a state with a large soft mass that does not couple to the Higgs boson~\cite{Dermisek:2016tzw}.
In this way, at the scale of vector-like quarks where stops are effectively generated, stops can get large soft masses without ever contributing in the RG evolution. If soft masses of vector-like quarks are comparable to the explicit vector-like mass terms the remaining threshold corrections to $m_{H_u}^2$ from the heavy sector can be small and the dominant contribution to  $m_{H_u}^2$ originates from two-loop effects (gluino and gauge contributions from heavy scalars).

\begin{figure}[t]
\begin{center}
\includegraphics[width=2.in]{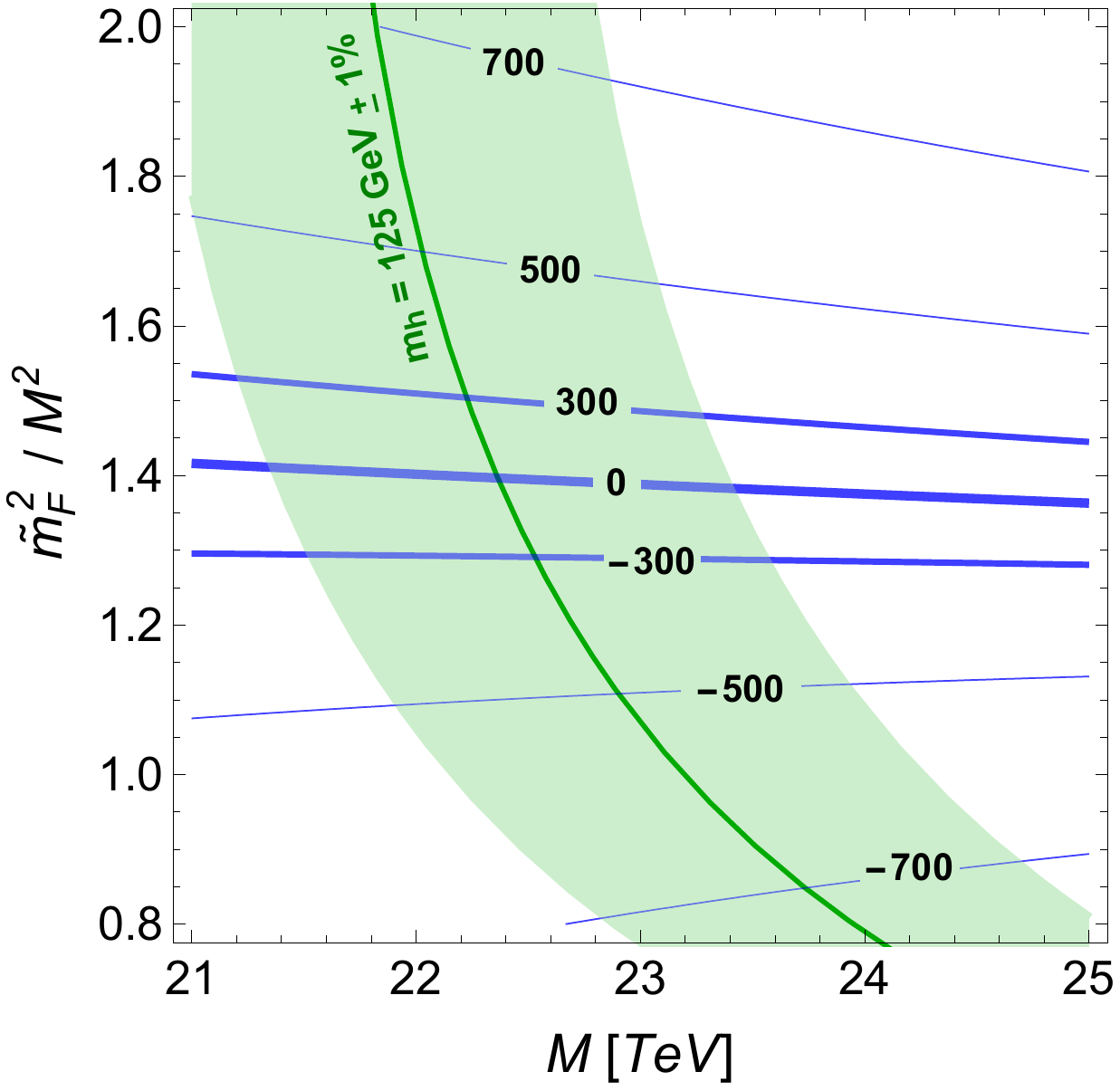}
\end{center}
\caption{Contours of constant contribution to $\tilde m_{H_u}^2/|\tilde m_{H_u}^2|^{1/2}$ [GeV]  from threshold corrections plotted  in  $\tilde m_F^2/M^2$ -- $ M$ plane.
Along the green line (and shaded area) $m_h = 125$ GeV ($\pm 1\%$)  in our approximation. The matching scale is $Q = m_{\tilde t_{1,2}}$ ($\simeq 9$ TeV in this case). 
}
\label{fig:RG}
\end{figure}

Assuming universal vector-like masses, $M$,  and soft masses, $\tilde m_F^2$, of vector-like quarks there are just two parameters that determine stop masses (and thus the Higgs boson mass). Contributions to   $m_{H_u}^2$ from threshold corrections  resulting from stop and heavy vector-like partners being somewhat spread in masses is plotted in Fig.~\ref{fig:RG}.
We can see that threshold corrections do not favor EWSB, they can be positive or negative, and there is $\sim 20\%$ range of soft mass squared that results in $< (300 \; {\rm GeV})^2$ correction to $m_{H_u}^2$. The plotted scenario assumes no mixing in the stop sector and thus requires $\sim 9$ TeV stop masses in order to generate sufficiently heavy Higgs boson mass. If sizable $A$-terms are also generated, the overall scale of the heavy sector can be lowered resulting in the correspondingly smaller size of threshold corrections to $m_{H_u}^2$.

Two-loop effects from gluino and gauge contributions from vector-like scalars both favor EWSB. These effects, as already mentioned for the gluino, still allow for a sizable gap between the mediation scale and the scale of verctor-like quarks.  Thus  the robust feature of radiatively triggered EWSB can be kept (at a loop suppressed level).

There is another interesting feature of MSSM extensions with vector-like families related to understanding relative strengths of gauge interactions. In the MSSM with one complete vector-like family, all three gauge couplings are asymptotically divergent. This means that starting with a large coupling at the GUT scale they run to fixed ratios in the IR. Far away from the GUT scale the ratios of gauge couplings are fixed and determined by the particle content. The preferred range of masses of vector-like quarks and leptons is a few TeV~\cite{Dermisek:2017ihj}. Similar findings apply to non-supersymmetric model with 3 vector-like families \cite{Dermisek:2012as, Dermisek:2012ke}, since the one loop beta function for $\alpha_3$ is identical in these two models and $\alpha_3$ is the main determining factor for the mass of vector-like families. This has nothing to do with limits on superpartner masses or the Higgs boson mass, but it independently points to multi-TeV scale.

\section{Little hierarchy from complexity}

Finally, I would like to discuss the possibility that a naturalness argument based on the largest individual contribution to given observable is too strong requirement. Such an argument tells us that if there is a parameter $A$ contributing to $X$ at $O(1)$ level (contribution of one parameter can be always normalized to 1), then in order to get $X=0.001$ there has to be something else that cancels this contribution. If there is just one more parameter $B$ contributing to $X$ then $B$ has to be carefully selected or tuned with 0.1\% precision to $A$. However, this direct relation between the size of the largest contribution and the level at which model parameters are tuned is there only in models with two parameters contributing to $X$.
If more than two parameters are contributing to a given observable, the ratio of the largest contribution to $X$ and the desired value of $X $ does not directly translate into how much the model parameters need to be tuned~\cite{Dermisek:2016zvl}. 

For example, if there are two more parameters contributing to $X$, $X= A-B-C-D$ (we label the contribution of parameter $A$ to $X$ by $A$ for notational simplicity), then there are endless possibilities to obtain $X=0.001$ and, among them, also such that no parameter needs to be specified beyond one digit.
For a randomly chosen $A$, for example $A=0.963$, the $B$ no longer needs to be 0.962, but it can be 0.9, with $C = 0.06$ and $D = 0.002$.  No matter what the remaining digits are, only the first digits of parameters need to be adjusted. This is just an example which illustrates the point, the contributions of various parameters do not need to be hierarchical. 

If there is no need to specify model parameters beyond one digit to obtain a given outcome then such an outcome is nothing special or tuned, it can be regarded as a completely ordinary possible result of the model. What is a tuned outcome in a model with 2 parameters may be a completely ordinary outcome in a more complex model.

Let us apply this approach to the CMSSM~\cite{Dermisek:2017xmd}. For simplicity, let us assume that model parameters are of the same order, 
$\mu,\; M_{1/2},\; -A_0,\; m_0 \simeq M_{SUSY}$,  
and let us vary them in $\pm 50\%$ range keeping only one digit specifying the departure from  $M_{SUSY}$, e.g.: 
$M_{1/2} = 0.8 M_{SUSY}$, $m_{0}^2 = 1.2 M_{SUSY}^2$ and so on.
These are completely ordinary input choices and therefore any outcome that does not depend on what the numbers beyond the first digit are can be regarded as completely ordinary.

\begin{figure}[t]
\begin{center}
\includegraphics[width=2.6in]{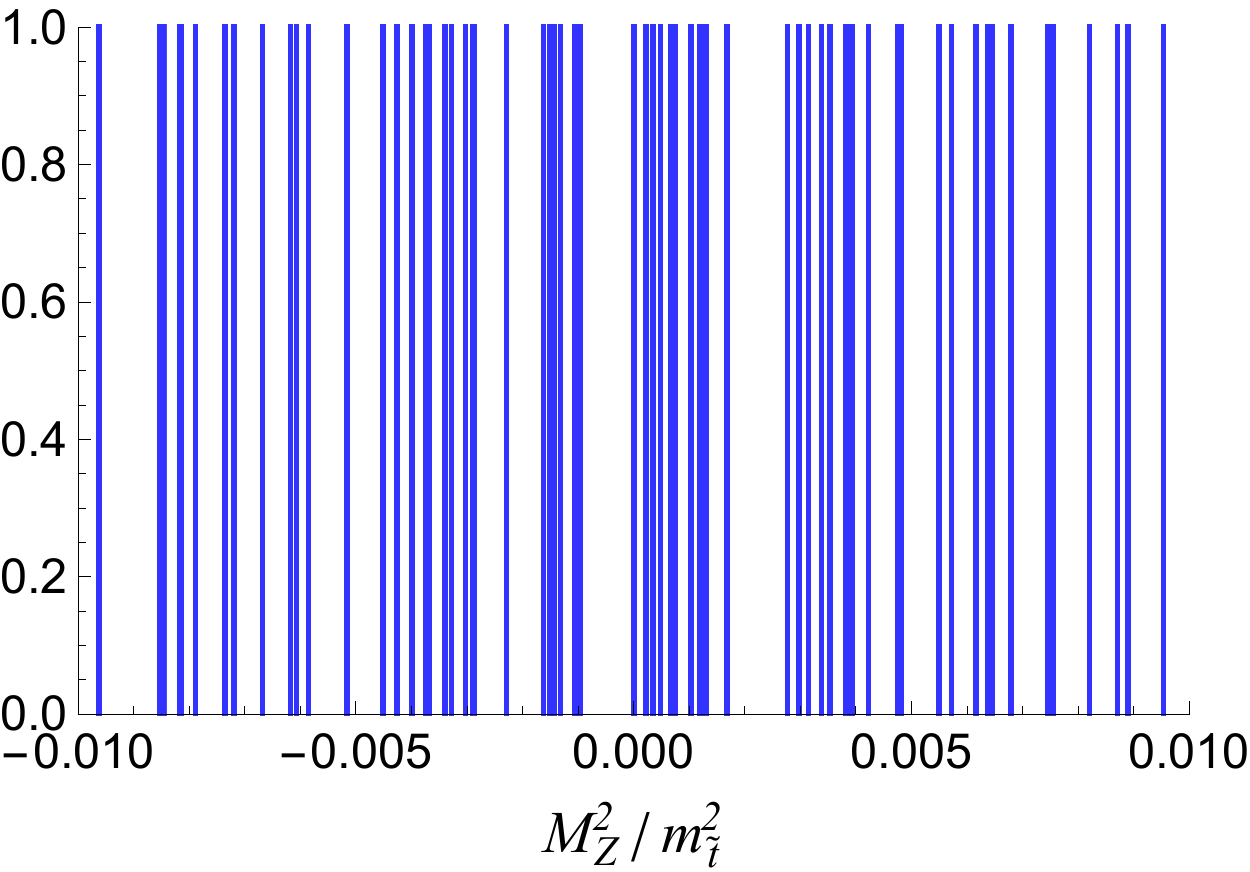}
\hspace{.5cm}
\includegraphics[width=2.6in]{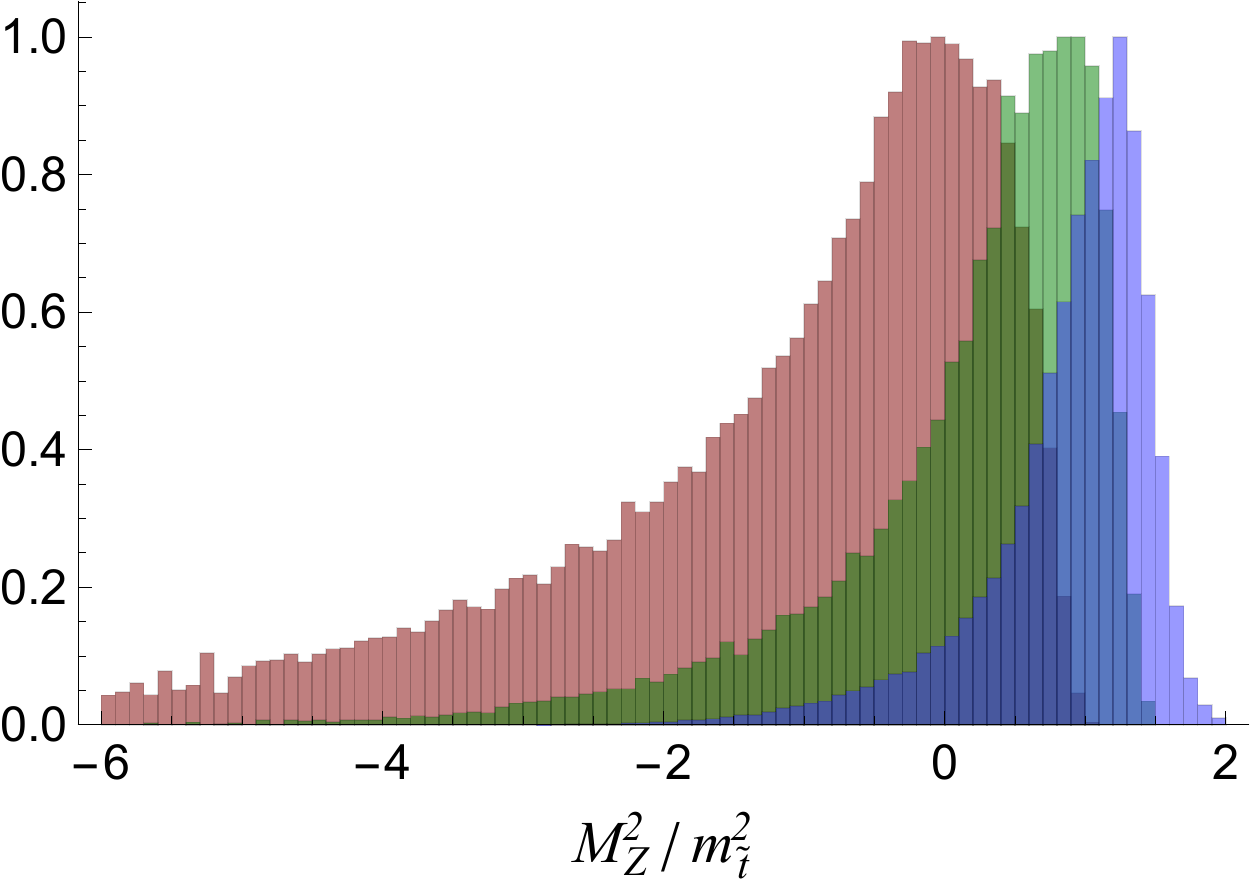}
\end{center}
\caption{Distribution of $M_Z^2/m_{\tilde t}^2$ in the CMSSM  for  $M_{SUSY} = 3$ TeV and $\tan\beta =10$ that corresponds to   $\mu$, $M_{1/2}$, $-A_0$ and $m_0^2$ being  generated in the 50\% interval around the central values set by $M_{SUSY}$ with their departures from $M_{SUSY}$ specified by one digit (blue). The left plot zooms in the region close to 0. The green and red distributions  correspond to shifted central value of $\mu$ to $\sqrt{2}M_{SUSY}$ and $2M_{SUSY}$. } 
\label{fig:mA0}
\end{figure}

The distributions of $M_Z^2/m_{\tilde t}^2$ obtained from such a scan, for  $M_{SUSY} = 3$ TeV and $\tan\beta =10$, is shown in Fig.~\ref{fig:mA0} (blue). The zoomed in distribution in the left plot shows the largest gap size between different outcomes to be $\sim 0.001$ (it varies little through the whole distribution). The largest gap in the distribution is an indicator of the smallest outcome that does not depend on specifying model parameters with more than 1 digit. Changing the central values of parameters or randomly choosing remaining digits shifts (in the first approximation) the whole distribution little to the left or to the right. Thus the outcomes smaller than the largest gap are accidental. However, an outcome as small as the largest gap is guaranteed to appear no matter what the remaining digits in all the parameters are (no matter how the distribution shifts)!

The outcome given by the largest gap can be viewed as a completely ordinary outcome resulting from model parameters being crudely selected (in this case by one digit or in 10\% intervals around random central value). The largest gap  $M_Z^2/m_{\tilde t}^2 \sim 0.001$ translates into $m_{\tilde t} \simeq 30 M_Z$.  The density of outcomes that  can be seen in overall distribution, Fig.~\ref{fig:mA0} (right), indicates what outcomes are slightly preferred. Note however, that the overall distribution can be changed when moving around the parameter space. 
For example, increasing the $\mu$ term favors smaller and smaller values as shown by green and red distributions in Fig.~\ref{fig:mA0}. Any value can be slightly favored in various regions. Nevertheless, the maximal gap size remains almost the same and thus the prediction for maximal hierarchy from minimally specified parameters is very robust.

\begin{figure}[t]
\begin{center}
\includegraphics[width=2.6in]{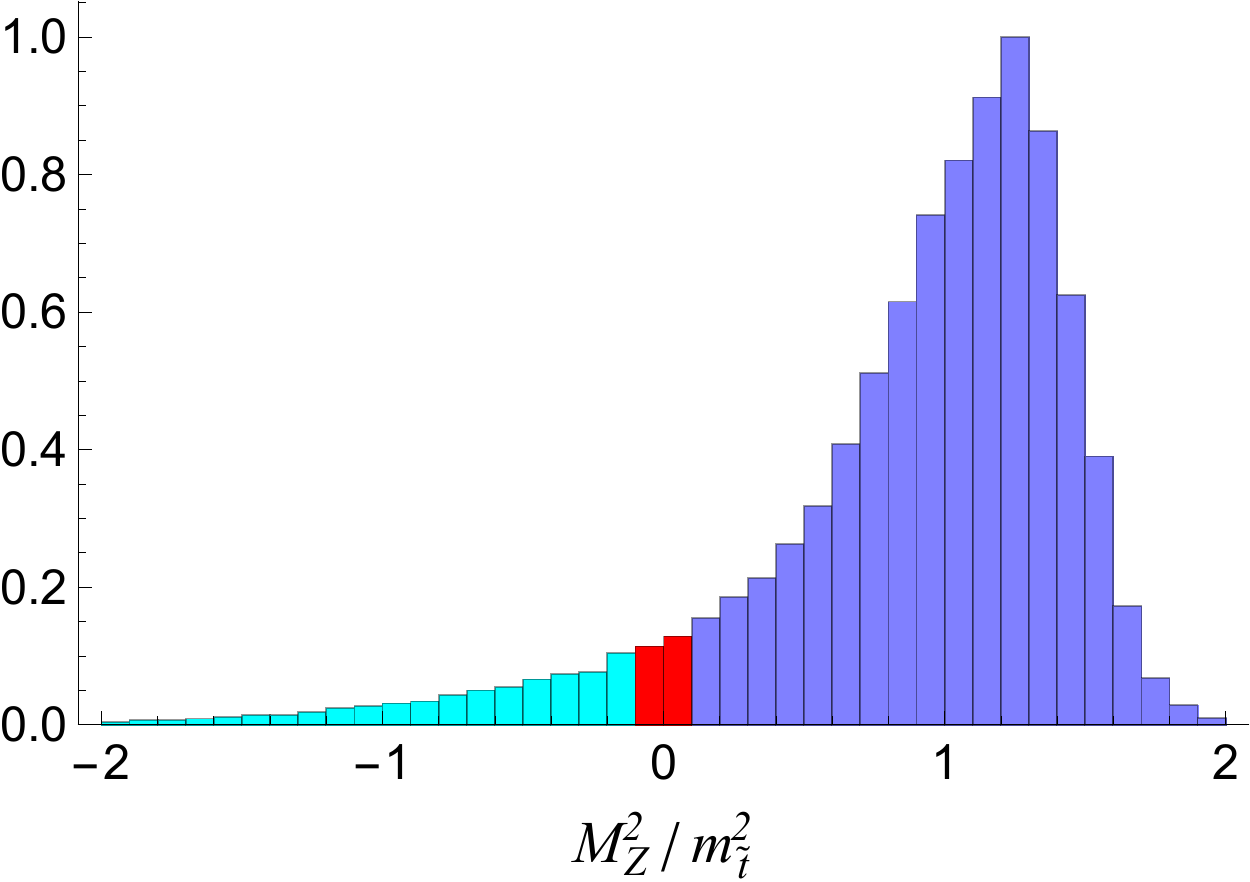}
\hspace{.5cm}
\includegraphics[width=2.in]{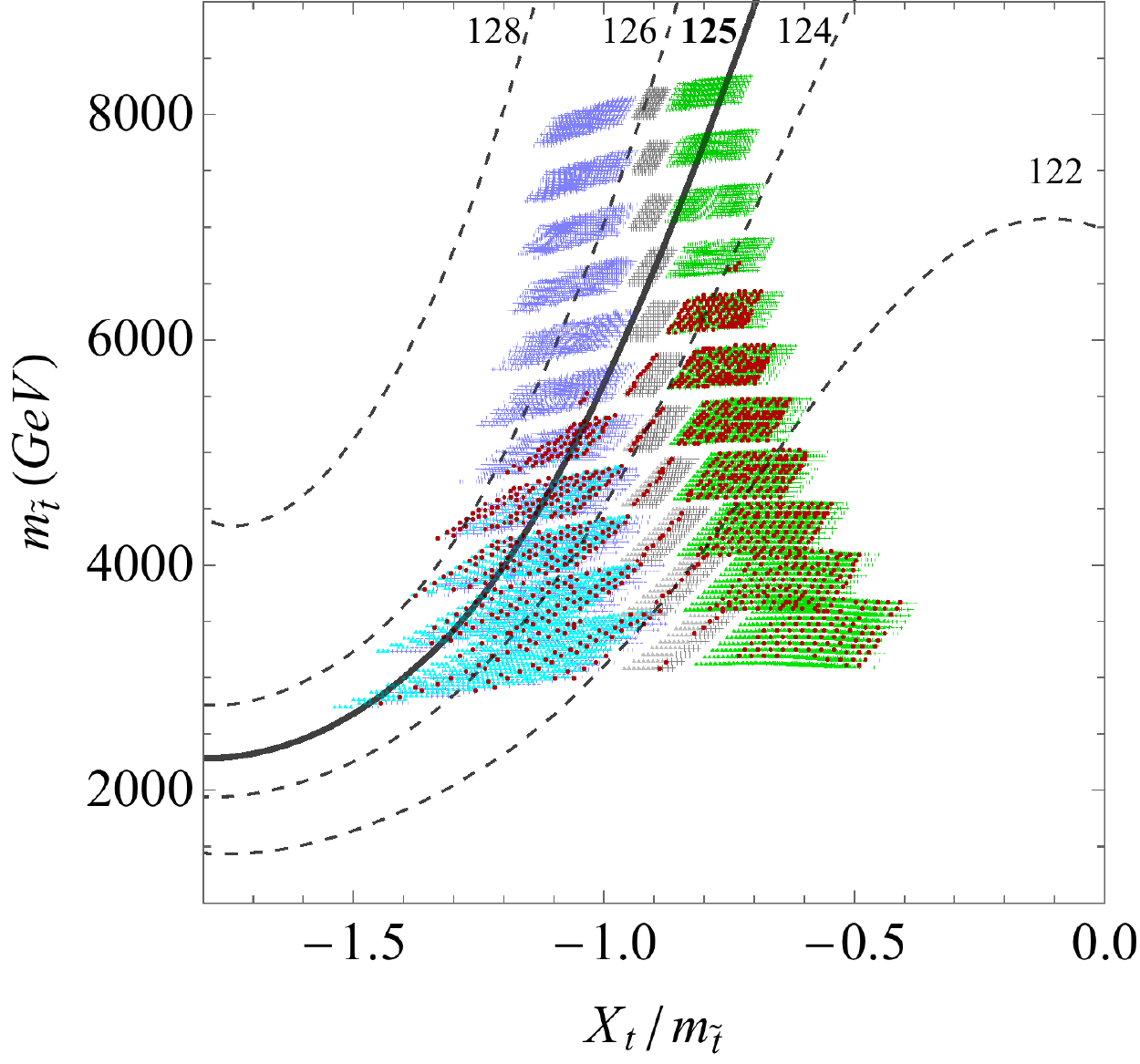}
\end{center}
\caption{Left: the blue distribution from Fig. 2
split into three colors that corresponds to color coding in the plot on the right. Right: CMSSM scenarios for  $M_{SUSY} = 3$ TeV and $\tan\beta =10$ generated in our procedure with negative $A_0$ (left, blue), $A_0 = 0$ (middle, gray) and positive $A_0$ (right, green) in the $m_{\tilde t} - X_t/m_{\tilde t}$ plane. Darker shade $+$ is used for points with $M_Z^2/m_{\tilde t}^2 > 0.1$, lighter shade triangle for  $M_Z^2/m_{\tilde t}^2 < -0.1$ and the points with $|M_Z^2/m_{\tilde t}^2|< 0.1$ in each distribution are highlighted (red) dots. Contours of constant Higgs boson mass are indicated for all the remaining parameters set to 3 TeV  at $m_{\tilde t}$ (gaugino masses are set to satisfy the GUT scale universality condition that would lead to $M_{\tilde g} = 3$ TeV).} 
\label{fig:higgs}
\end{figure}

The remaining question is whether the maximal hierarchy from minimally specified parameters is sufficient for the Higgs mass. The generated points corresponding to the blue distribution in Fig.~\ref{fig:mA0} are shown in the  $m_{\tilde t} - X_t/m_{\tilde t}$ plane in Fig.~\ref{fig:higgs}. For completeness, similar distributions for  zero $A$ terms and positive $A$ terms are also shown. The blue points from the original distribution are split into three groups: red near 0, darker blue above and lighter blue below, just to see where they fall. Any point can be made realistic by adjusting the $\mu$ term. Since the maximal possible hierarchy from model parameters specified with one digit suggests $m_{\tilde t} \lesssim 3$ TeV, we see that only the scenarios with negative $A$ terms can have sufficiently heavy stops to explain the Higgs boson mass.

\section{Summary}

It became clear that the MSSM with all superpartners at the EW scale, as envisioned based on a simple naturalness argument, is not realized in nature. Either the SUSY model is different than we thought, or our perception of naturalness must change, or both. Whenever there is a puzzling situation, there is also  a great opportunity for both theory and experiment.

\end{document}